\documentclass[amsmath,amssymb,aps,pra,reprint,twocolumn]{revtex4-2}
\usepackage[centering,hmargin=0.71in,tmargin=0.7in,bmargin=1.07in]{geometry}

\usepackage{graphicx}
\usepackage{orcidlink}
\usepackage{braket}
\usepackage{bm}
\usepackage{paralist}

\DeclareMathOperator{\Tr}{Tr}
\newcommand{\ic}{\mathrm{i}\mkern1mu} 
\newcommand{\ketbra}[3][]{\ket{#2}_{#1}\hspace{-0.2em}\bra{#3}}
\newcommand{\PMIS}{P_\text{MIS}}
\newcommand{\HP}{\mathbb{H}}
\newcommand{\A}{\mathcal{A}}
\newcommand{\V}{\hat{V}}
\newcommand{\X}{\hat{X}}
\newcommand{\Y}{\hat{Y}}

\newcommand{\n}{\hat{n}}
\newcommand{\taui}{\tau_{\text{i}}}
\newcommand{\tauf}{\tau_{\text{f}}}
\newcommand{\taum}{\tau_{\text{m}}}
\newcommand{\tad}{t_{\text{ad}}}
\newcommand{\ti}{t_{\text{i}}}
\newcommand{\tf}{t_{\text{f}}}
\newcommand{\tmax}{t_{\text{max}}}
\newcommand{\Rb}{R_{\text{b}}}
\newcommand{\Rs}{R_{\text{s}}}
\newcommand{\W}{\Omega}
\newcommand{\Wi}{\Omega_{\text{i}}}
\newcommand{\Wf}{\Omega_{\text{f}}}
\newcommand{\Wmax}{\Omega_{\max}}
\newcommand{\D}{\Delta}
\newcommand{\Di}{\Delta_{\text{i}}}
\newcommand{\Df}{\Delta_{\text{f}}}
\newcommand{\Dm}{\Delta_{\text{m}}}
\newcommand{\DLB}{\Delta_{\text{LB}}}
\newcommand{\DUB}{\Delta_{\text{UB}}}
\newcommand{\um}{{\textmu}m}
\newcommand{\us}{{\textmu}s}

\begin{document}


\title{Hardness-dependent quantum adiabatic schedules for the maximum-independent-set problem}

\author{S\'ebastien Perseguers\,\orcidlink{0000-0003-1022-8183}}
\email[Contact author:\ ]{contact@gradiom.ch}
\affiliation{Gradiom S\`arl, Avenue de Tivoli 4, 1700 Fribourg, Switzerland}

\date{June 10, 2025}

\begin{abstract}
	We propose a numerical approach to design highly efficient adiabatic schedules for analog quantum
 computing, focusing on the maximum-independent-set problem and neutral atom platforms. On the basis
 of a representative dataset of small graphs, we present numerical evidence that the optimum schedules
 depend principally on the hardness of the problem rather than on its size. These schedules perform better
 than the benchmark protocols and admit a straightforward implementation in the hardware. This allows
 us to extrapolate the results to larger graphs and to successfully solve moderately hard problems using
 QuEra’s 256-qubit Aquila computer. We believe that extending our approach to hybrid algorithms could
 be the key to solving the hardest instances with the current technology, making yet another step toward
 real-world applications.
\end{abstract}


\maketitle


\section{Introduction}

In recent years, quantum computers have emerged as a promising way to solve complex optimization tasks in science and industry, notably those based on combinatorics where an exhaustive search is not tractable~\cite{Bova2021,
Halffmann2023, Abbas2024}. Typically, the problem of interest is encoded in the ground state of a quantum Hamiltonian, which can be reached through adiabatic evolution starting from an easy-to-prepare initial state~\cite{
Fahri2001, Lucas2014, Albash2018, Grant2020}. Advanced protocols have to be developed to overcome not only the hardware limitations, such as qubit decoherence and control errors, but also the intrinsic complication of a vanishing gap in the eigenspectrum during the adiabatic evolution~\cite{
DelCampo2013, Ohkuwa2018, Crosson2021, Cirac2022}.

Here we focus on the maximum-independent-set (MIS) problem, where the goal is to find the largest set of vertices that do not share an edge in a graph. Such problems appear naturally in practical applications, for instance, network design~\cite{Wurtz2024} or financial analysis~\cite{Boginski2005}, and are known to be NP-hard in general~\cite{Karp1972}. Finding a MIS is still a difficult task in the particular case of unit disk graphs, in which any two vertices are connected by an edge if they are closer than a fixed distance~\cite{Clark1990, Andrist2023}. This family of graphs admits a direct mapping to arrays of Rydberg atoms, which explains the popularity of the MIS problem in analog quantum computing~\cite{Pichler2018, Henriet2020, Ebadi2022,
Cain2023, Dalyac2023, Schiffer2024, Schuetz2024}. In that setting, optical tweezers enable arbitrary arrangements of atoms and controlled laser pulses manipulate the quantum states~\cite{Lukin2001, Wurtz2023}. The lasers then drive the evolution of the system according to schedules that are specific to the problem to be solved.

The precise description of the schedules is of utmost importance since it conditions the success of the quantum algorithm. Many methods based on adiabatic criteria or on carefully chosen ansatzes can be applied in this case: quantum approximate optimization algorithm and variational quantum eigensolver~\cite{Farhi2014, Moll2018, Cerezo2021}, shortcut to adiabadicity and counterdiabatic (CD) protocol~\cite{Demirplak2003, Berry2009, Sels2017,
Zhang2024}, Bayesian and recursive optimization~\cite{Finzgar2024,
Finzgar2024b}, machine learning~\cite{Coelho2022}, and variational sampler~\cite{Wurtz2024b}, to name a few. Most of the aforementioned methods require multiple iterations to be run on a quantum device and to be updated by a classical optimizer and thus belong to the class of hybrid algorithms~\cite{Ge2022}. By their very nature, these algorithms cost both time and money as they make intense use of the quantum hardware, which motivates us to search for a more direct way of designing efficient schedules.

A detailed analysis of the classical dynamics of the simulated annealing for the MIS problem reveals that the graphs with many suboptimal solutions are likely to trap the algorithm in local minima, hence missing the global solution~\cite{
Ebadi2022}. This observation leads to the definition of the hardness parameter
$\HP = D_{|\text{MIS}|-1} / (|\text{MIS}| \, D_{|\text{MIS}|})$, where $D_k$
is the degeneracy of the independent sets of size $k$. On the basis of the
theory of Markov chains, $\HP$ is shown to determine the performance of simulated annealing as it provides a lower bound on the inverse of its spectral gap and, by extension, on the hitting time for finding the solution. By analogy with the adiabatic theorem~\cite{Albash2018}, this suggests that the hardness parameter may play a key role during the optimization of the schedules in the quantum setting, which is the starting point of our study.

In what follows, we show how a precomputation of optimized schedules for small graphs can lead to highly efficient protocols for graphs of larger size on the quantum
device~\cite{note:order}. First, we review the MIS encoding in Rydberg atoms, emphasizing the importance of well-chosen boundary conditions. Then, in Sec.~\ref{sec:numerics} we describe how to optimize the schedules on the basis of numerical simulations of small graphs. Finally, in Sec.~\ref{sec:demonstrations} we present the results of cloud quantum computing demonstrations. Detailed information and calculations are provided in the appendices.

\subsection{{MIS} encoding in Rydberg atom arrays}

The analog quantum dynamics of $N$ atoms is governed by the Rydberg
Hamiltonian $H = H_{\text{drive}} + H_{\text{cost}}$, where the driving
part and the cost function are~\cite{Ebadi2022}
\begin{subequations}
	\label{eq:H}
	\begin{align}
		\frac{H_{\text{drive}}(t)}{\hbar} &= 
			\frac{\W(t)}{2} \sum_{i=1}^N e^{\ic\phi(t)}\ketbra[i]{0}{1}
			+ \text{H.c.},\\
		\frac{H_{\text{cost}}(t)}{\hbar} &= 
			-\D(t) \sum_{i=1}^N \n_i + \sum_{i<j}V_{ij}\n_i\n_j.
	\end{align}
\end{subequations}
In these equations, $\W$ is the Rabi drive amplitude, $\phi$ its phase, $\D$
the detuning, and $\n_i = \ketbra[i]{1}{1}$ detects the Rydberg
excitation of the $i$th atom.
The two-body potential is $V_{ij} = C_6/|\textbf{r}_i-\textbf{r}_j|^6$,
with $C_6$ a constant for the van der Waals interaction between two Rydberg
states at positions $\textbf{r}_i$ and $\textbf{r}_j$. Because of the large
decay exponent, this interaction prevents the presence of any doubly
excited state if two atoms are close together but has a negligible impact
if they are further apart. This enables the encoding of MIS problems on unit
disk graphs, where the so-called blockade radius of the Rydberg interaction
corresponds to the maximum distance between any two connected vertices of the
graph, as illustrated in Refs.~\cite{Andrist2023, BloqadeSubspace2024};
see Appendix~\ref{sec:MIS} for the mathematical description of this encoding.

The boundary conditions at initial and final times $\ti$ and $\tf$ of
the adiabatic quantum computation are usually stated as follows:
\begin{subequations}
	\label{eq:bc}
	\begin{gather}
		\Wi = \Wf = 0,\\
		\Di < 0 < \Df.
	\end{gather}
\end{subequations}
In practice, we implement the first inequality as $\Di \leq -\delta_\D$,
where $\delta_\D$ is a small positive constant that incorporates the main sources
of error in the quantum hardware, so that it is ensured that all atoms start
in the easy-to-prepare ground state $\ket{0}$. Regarding the MIS encoding,
the second inequality turns out to be necessary but not sufficient,
as the final detuning has to lie in a narrow interval that depends on the graph
geometry, see Appendix~\ref{sec:MIS_bounds}. For instance, the lower and upper
bounds of the detuning for unit disk graphs on a square lattice are given by
\begin{equation}
	\DLB = \frac{V_0}{12} \leq \Df \leq \frac{V_0}{8} = \DUB,
\end{equation}
where $V_0 = C_6/a^6$ is the interaction strength between adjacent nodes
separated by distance $a$. Note that such bounds are not universal,
but they are satisfied by the vast majority of graphs for a given lattice.
\newline

\subsection{Quantum hardware}

Throughout this paper, we consider QuEra's Aquila device, whose characteristics
are described in detail in Ref.~\cite{Wurtz2023} and which is easily accessible
through the \texttt{AMAZON BRAKET} interface~\cite{AWS2024}. Of course, all
results extend to other devices and frameworks such as Pasqal's \texttt{PULSER}
and also to other types of lattices (e.g., triangular)~\cite{Silverio2022,
Leclerc2024}. The interaction coefficient $C_6$ between two excited states in
Aquila is $5\,420\,503\text{\,MHz}\text{\,\um}^6$, and two limitations are
particularly important in our study: first, the maximum duration of the
protocol $\tmax = 4$\,\us\ to ensure a coherent evolution, and second, the
maximum Rabi frequency $\Wmax = 15.8$\,MHz due to the limited laser power for
driving the ground-Rydberg transition. We ignore the hardware errors in the
simulations, but they could be taken into account to fine-tune the results. The
only error that we consider here is the noise $\delta_\Delta \approx 1$\,MHz in
the global detuning since it plays a role in the upper bound of $\Di$. Unless
otherwise specified, all units are micrometers, microseconds, and megahertz in
what follows.

In Aquila, the Rabi amplitude and the detuning must be defined as piecewise linear and not smooth functions, while the phase must be piecewise constant. However, the step size can be as small as 0.05\,\us, which results in 80 intervals for the longest coherent evolution. The schedules are therefore well approximated in the quantum hardware, but one could slightly improve the results of the demonstrations by taking into account this hardware capability in the numerical simulations.

\section{Numerical optimization}
\label{sec:numerics}

We consider a fixed set of 500 representative graphs of size 8 to 17, which
allows us to perform fast numerical simulations and to use efficient
deterministic optimization algorithms, see Appendix~\ref{sec:datasets}. This
dataset is such that the correlation between the size of the graphs and their
hardness parameter is minimized. Therefore, the results that depend principally
on $\HP$ can be extrapolated, to a certain extent, to graphs of any size.

In this work, we focus on the probability to reach a MIS solution at the end of
the adiabatic evolution, which we denote by $\PMIS$ and use as the score
function to be optimized for some parametrization $\Theta$ of the schedules:
\begin{equation}
	\PMIS(\Theta) = \sum_{i = 1}^{D_{|\text{MIS}|}}
	\big|\,\braket{\psi(\Theta,\tf)\,|\,\text{MIS}_i}\,\big|^2,
	\label{eq:PMIS}
\end{equation}
where $\ket{\psi(\Theta,\tf)}$ is the final quantum state and
$\ket{\text{MIS}_i}$ is the basis vector that corresponds to the $i$th solution
of the MIS problem. Note that most studies consider the \textit{approximation
ratio} instead (normalized mean energy of the final state), but in our opinion
only the lowest energy level and not the full energy spectrum has practical
utility because of the very definition of the hardness parameter.

Following the ``best practices'' listed in Ref.~\cite{Wurtz2023}, we consider
protocols for which the Rabi amplitude $\Omega(t)$ reaches $\Wmax$ during the
evolution, so that the associated dynamic blockade radius $\Rb$ becomes
$(C_6/\Wmax)^{1/6} \approx 8.4$. In order to generate a unit disk graph on the
nearest and next-nearest neighbors of a square lattice, this radius should
satisfy $\sqrt{2} a < \Rb < 2 a$, and consequently $4.2 < a < 5.9$. We set
$a = 5$, which is close to the geometric mean of these bounds and which yields
good results in preliminary calculations. We consider a fixed evolution time
$\tf = 1$ so that $\PMIS$ varies as much as possible depending on the various
protocols; much shorter or longer times lead to either extremely bad or
extremely good results independently of the precise shape of the schedules.

In the following sections, we optimize and compare two paradigmatic
protocols based on either piecewise linear or counterdiabatic schedules.

\subsection{Piecewise linear protocol}
\label{sec:numerics_linear}

\begin{figure}[t!]
	\includegraphics[width=0.953\columnwidth]{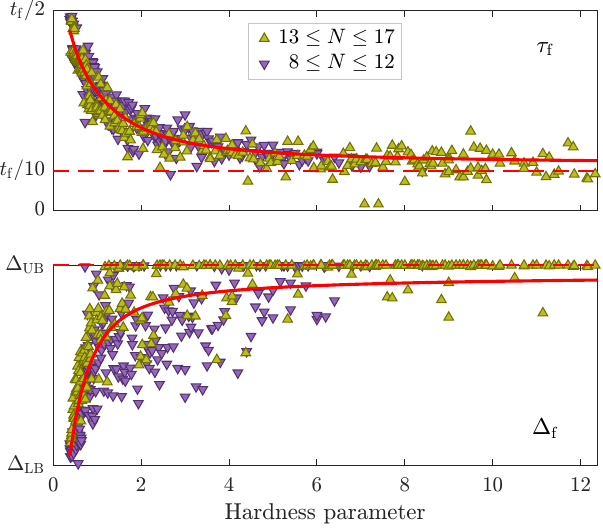}
	\caption{Optimum parameters of the linear schedules for each graph of the
	dataset. These values depend on $\HP$ and converge to a constant (dashed
	red lines) for sufficiently hard graphs. This behavior is particularly
	strong for $\tauf$ (top) and $\Df$ (bottom) at the end of the adiabatic
	evolution, with no statistical difference between the	smallest graphs
	(purple triangles) and the largest ones (yellow-green triangles). We fit
	all points to extract a smooth and monotonic dependence on $\HP$ (thick
	red lines).}
	\label{fig:optim_params_Lin4_HP}
\end{figure}

The typical adiabatic protocol is based on piecewise linear schedules with
two breakpoints,
where $\W$ is slowly ramped up, then $\D$ is ramped from negative to
positive, and finally $\Omega$ is ramped down; see Fig.~3.1b in Ref.~\cite{Wurtz2023}
or Fig.~2c in Ref.~\cite{Kim2024}. Usually, these schedules are
symmetric in the sense that the initial and final ramp times $\taui$ and
$\tauf$ are equal, and that the initial and final detunings $\Di$ and $\Df$
have opposite values. In our case, we are looking for schedules that are as
general as possible, so we keep these parameters independent. Moreover,
we do not impose the condition that the breakpoints of $\W$ coincide with those of $\D$.

\begin{figure}[h]
	\includegraphics[width=0.972\columnwidth]{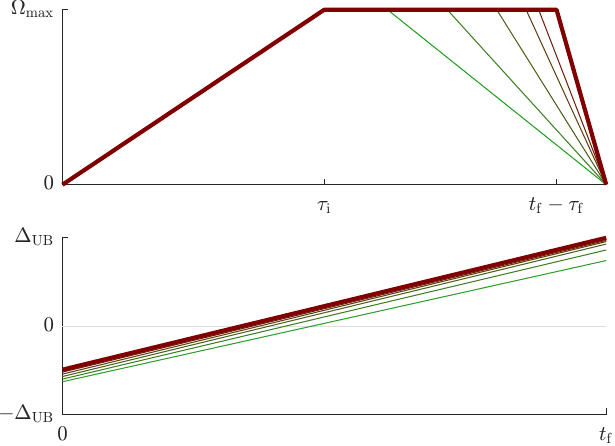}
	\caption{Optimized piecewise linear schedules with four free variables. The
	Rabi amplitude (top) and the detuning (bottom) depend on the hardness	of the
	graph and are illustrated here for $\mathcal{HP} \in \{0.5, 1, 2, 4, 8\}$
	(thin lines, from green to maroon) and in the limit $\HP \to \infty$ (thick
	dark maroon line). In this limit, the optimized values become
	$\taui \approx \tf/2$, $\tauf \approx \tf/10$,	$\Di \approx -\DUB/2$, and
	$\Df = \DUB$.}
	\label{fig:optim_schedules_Lin4_HP}
\end{figure}

After a few optimization steps, it appears that the flat portions of the
detuning tend to vanish; we therefore consider a simple linear function for
this schedule. Then we optimize the four remaining parameters
$(\taui, \tauf, \Di, \Df)$ for each graph of the dataset. A noticeable
observation is that the optimum parameters depend mostly on the hardness
parameter and not on the size of the graph. Consequently, we can fit these data
to generate a model of optimized schedules, as illustrated in
Figs.~\ref{fig:optim_params_Lin4_HP} and~\ref{fig:optim_schedules_Lin4_HP}.

\begin{figure}[h]
	\includegraphics[width=0.972\columnwidth]{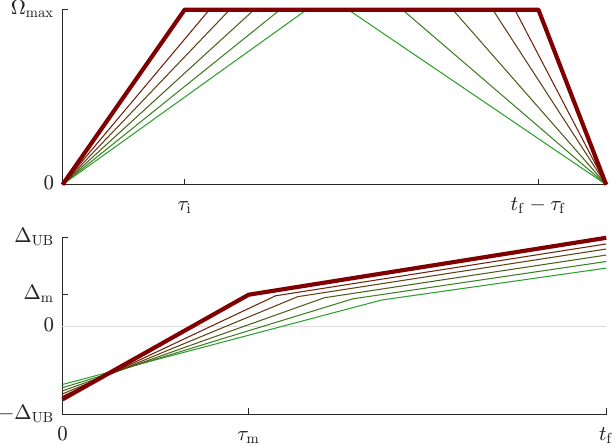}
	\caption{Optimized piecewise linear schedules with six free variables.
	In this case, $\taui$ displays a clear dependence on $\HP$. In the limit of
	infinite hardness, the optimized values are $\taui \approx 0.22 \tf$ and
	$\tauf \approx 0.12 \tf$ for the	Rabi amplitude (top) and
	$\Di \approx -0.83 \DUB$, $\taum \approx 0.34 \tf$,
	$\Dm \approx 0.36 \DUB$, and $\Df = \DUB$ for the detuning (bottom).}
	\label{fig:optim_schedules_Lin6_HP}
\end{figure}

The hardness-dependent linear schedules yield very good results compared with
existing protocols, which motivates us to go one step further by considering
two more degrees of freedom $(\taum,\Dm)$ for the detuning; see
Fig.~\ref{fig:optim_schedules_Lin6_HP}. The resulting optimized schedules are
still easily implementable, yet they perform much better than all benchmark
protocols; see Sec.~\ref{sec:numerics_results}.

\begin{figure}[t!]
	\includegraphics[width=\columnwidth]{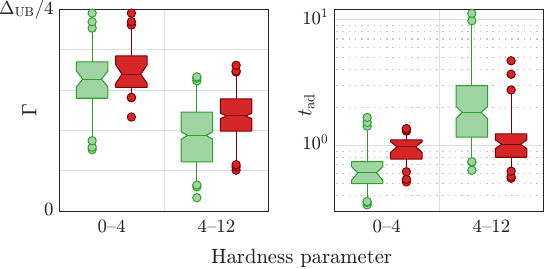}
	\caption{Adiabatic criteria for the $\HP$-linear protocol with six parameters.
	Left:~Minimum energy gap for ``easy'' ($\HP \leq 4$) and ``hard'' ($\HP > 4$)
	graphs, 	based on the schedules illustrated in Fig.~\Ref{fig:optim_schedules_Lin6_HP}
	with $\HP = 1$ (light green) and $\HP \to \infty$ (maroon). The gap by itself
	does not explain the hardness dependence of the optimum schedules, as shorter
	ramp times of the detuning (maroon boxes) are slightly beneficial for all
	graphs. Right:~Corresponding adiabatic timescales. In this case, $\tad$ is
	indeed minimized if one applies the schedules according to the hardness of
	the graphs (leftmost and rightmost boxes).}
	\label{fig:adiabatic_Lin6_HP}
\end{figure}

Let us now provide some hints about the performance of the hardness-dependent
linear schedules. The adiabatic condition for quantum computing is
traditionally given by~\cite{Albash2018}
\begin{equation}
	\tf \gg \max_{s \in [0,1]}
	\frac{|\bra{\varepsilon_1(s)} \partial_s H(s) \ket{\varepsilon_0(s)}|}
	{|\varepsilon_1(s) - \varepsilon_0(s)|^2} = \tad,
	\label{eq:tLB}
\end{equation}
where $s=t/\tf$ is the dimensionless time, $\ket{\varepsilon_0(s)}$ is the
instantaneous ground state of $H(s)$ with energy $\varepsilon_0(s)$, and
$\ket{\varepsilon_1(s)}$ is any of the first excited states; for simplicity,
we consider only the nondegenerate MIS instances here. In particular,
Eq.~(\ref{eq:tLB}) implies that the minimum energy gap
$\Gamma = \min_s \varepsilon_1(s) - \varepsilon_0(s)$ should be large to
maintain adiabaticity. Unsurprisingly, this gap gets smaller as the hardness
parameter increases: from the numerical simulations of the linear schedules
with six parameters, we find $\Gamma \propto \exp(-\alpha \HP^\beta)$ with
$\alpha \approx 0.76$ and $\beta \approx 0.36$. In this respect, it seems
counterintuitive that the Rabi amplitude should be ramped up and down more
rapidly for harder graphs; see Fig.~\ref{fig:optim_schedules_Lin6_HP}. Faster
variations are likely to introduce stronger diabatic transitions which, in this
case, would not be compensated by a larger gap to satisfy the adiabatic
condition. However, we show in Fig.~\ref{fig:adiabatic_Lin6_HP} that the
optimized schedules do minimize the timescale $\tad$ on average, which confirms
that adiabaticity lies at the heart of the protocol's performance.

\subsection{Counterdiabatic protocol}
\label{sec:numerics_cd}

The limited coherence time of the quantum state in the hardware is a major
source of error in adiabatic computing. Well-established methods to overcome
such limitations include the shortcuts to adiabaticity, among which is the CD
driving~\cite{Odelin2019}. The idea of the CD driving is to add a gauge
potential to the adiabatic Hamiltonian such that the transitions between
eigenstates are suppressed during the evolution. However, the calculation of
the exact gauge potential is very difficult in general, and these mathematical
terms may not be always adapted to an implementation in the quantum hardware.
Nevertheless, by ignoring the interactions of the Rydberg Hamiltonian, one is
able to find a suitable gauge potential that utilizes the phase of the Rabi
drive~\cite{Zhang2024}. In this study, we propose to generalize the adiabatic
schedules of this so-called analog counterdiabatic quantum computing (ACQC)
protocol as follows:
\begin{subequations}
	\label{eq:smooth_functions}
	\begin{align}
		\W_\text{ad}(t) &= \Wmax\, \sin\left( \frac{\pi}{2}\, \sin\left(\theta_\W(t)\right)\right)^2,\\
		\D_\text{ad}(t) &= \frac{\Di+\Df}{2} + \frac{\Di-\Df}{2}\cos\left(\theta_\D(t)\right),
	\end{align}
\end{subequations}
where $\theta(t)$ is any monotonic smooth function that ranges from $0$ to
$\pi$; here we set $\theta_{\W,\D}(t) = \pi t/\tf$. Moreover, on the basis of the
variational procedure described in Ref.~\cite{Sels2017}, we improve on the ACQC
protocol by including the Rydberg interactions in the calculation of the
CD terms, thus adapting the gauge potential to each different graph; see
Appendix~\ref{sec:CD}. The detuning is not altered by this operation, whereas
the Rabi drive becomes
\begin{subequations}
\begin{align}
	\W &= \sqrt{\W_\text{ad}^2 + \W_\text{cd}^2},\\
	\phi &= \arctan(\W_\text{cd}/\W_\text{ad}),
\end{align}
\end{subequations}
with
\begin{equation}
	\W_\text{cd} = \frac{\W_\text{ad}\dot{\D}_\text{ad}
	- \dot{\W}_\text{ad}\D_\text{ad} + \dot{\W}_\text{ad}\,T_1\,\nu}
	{\W_\text{ad}^2 + \D_\text{ad}^2 - 2\D_\text{ad}\,T_1\,\nu + T_2\,\nu^2}
\end{equation}
and where $0 \leq \nu \leq V_0$ is an additional free parameter that
interpolates between the standard gauge potential and the generalized gauge
potential; see Appendix~\ref{sec:CD_variational}. In the
latter equation, the variables $T_1$ and $T_2$ depend on the interaction terms
$V_{ij}$ and are therefore specific to the graph to be solved. As expected, the
CD contribution vanishes for $\tf \to \infty$ since both $\dot{\W}_\text{ad}$
and $\dot{\D}_\text{ad}$ tend to zero in this limit.

\begin{figure}[h!]
	\includegraphics[width=0.972\columnwidth]{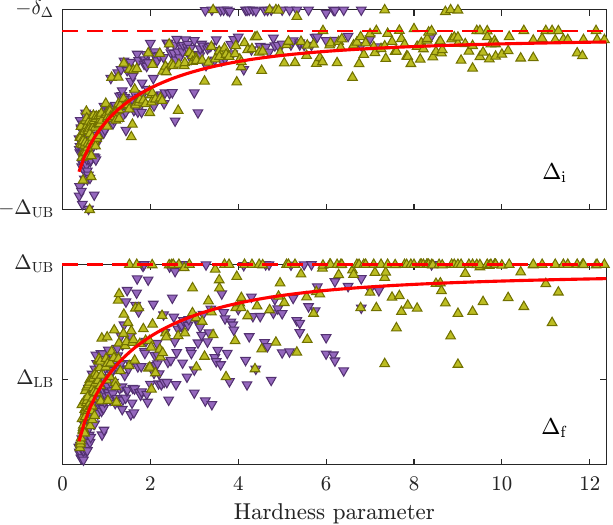}
	\caption{Optimum parameters for the CD schedules. The initial detuning (top)
	reaches the upper bound $-\delta_\Delta$ for several graphs but tends to a
	lower value $\Di \approx -\DUB/10$ for the hardest graphs. The final detuning
	(bottom) also increases with the hardness parameter and quickly reaches the
	limit $\DUB$. The coefficient $\nu$ (not shown) displays a similar behavior
	and	tends to approximately $V_0/80$ in the current setting.}
	\label{fig:optim_params_ACQC3_HP}
\end{figure}

\begin{figure}[h!]
	\includegraphics[width=0.972\columnwidth]{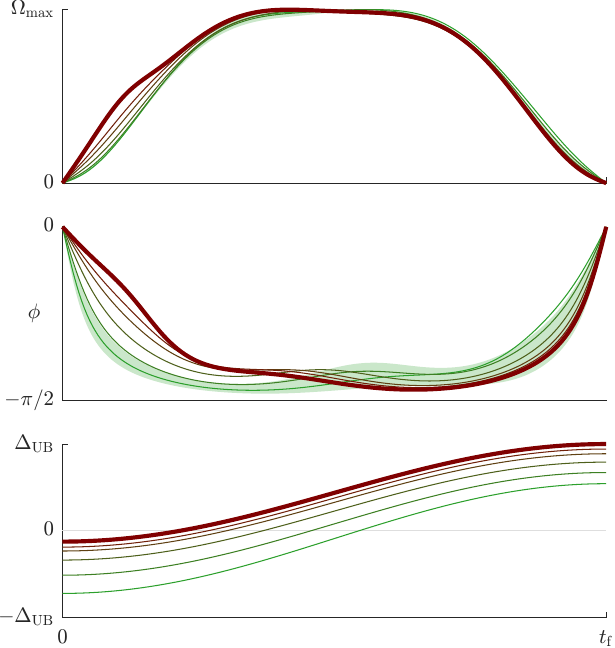}
	\caption{Optimized counterdiabatic schedules with three free variables. Unlike
	the detuning (bottom), the Rabi amplitude (top) and phase (middle) depend not
	only on $\HP$ but also on $\tf$ and	on the graph-dependent variables $T_1$
	and $T_2$. Here the curves are displayed for $\tf = 1$ and for the mean
	values of the dataset $\overline{T}_1 \approx 0.93$ and
	$\overline{T}_2 \approx 1.46$. The light-green area represents the
	variation solely due to $T_1$ and $T_2$ for $\HP = 0.5$.}
	\label{fig:optim_schedules_ACQC3_HP}
\end{figure}

Similarly as for the piecewise linear protocol, we optimize the schedule
parameters $(\Di,\Df,\nu)$ for all graphs of the dataset; see
Fig.~\ref{fig:optim_params_ACQC3_HP}. The interpolating coefficient $\nu$
tends to a strictly positive value for the hardest graphs, indicating that
the generalized gauge potential plays an important role in the CD protocol.
Note that the easy graphs exhibit optimized values of $\Df$ that are smaller
than the statistically determined constant $\DLB$. This is because the
exact lower bounds on the detuning are much smaller in that
case; see Fig.~\ref{fig:bounds_analysis} in Appendix~\ref{sec:MIS_bounds}.
The shape of the CD phase is highly nontrivial, and
we observe a strong shift toward higher values of the detuning for hard MIS
instances; see Fig.~\ref{fig:optim_schedules_ACQC3_HP}. Finally, we also tried
to further improve the results by considering more evolved $\theta$ functions\textemdash
in particular, to mimic the skewed shape of the Rabi amplitude
in Fig.~\ref{fig:optim_schedules_Lin4_HP}\textemdash
but no significant gain was found in this case.
\newline

\subsection{Results}
\label{sec:numerics_results}

\begin{figure}[t]
	\includegraphics[width=0.972\columnwidth]{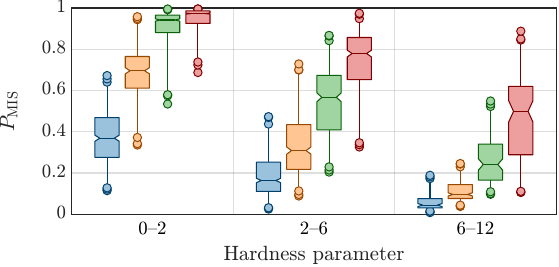}
	\caption{Box plot of the probability of reaching the MIS depending on the
	hardness of the 500 graphs of the dataset. The piecewise linear (blue) and ACQC
	(orange) protocols serve as benchmarks. For hard graphs, the generalized
	$\HP$-ACQC (green) and $\HP$-linear (red) protocols developed in this study
	lead to an average success probability that is increased by approximately 150\%
	and 1\,000\%, respectively.}
	\label{fig:optim_PMIS_box}
\end{figure}

In Fig.~\ref{fig:optim_PMIS_box}, we compare the probability $\PMIS$ to reach the
solution for four protocols: the piecewise linear and ACQC protocols defined
in Ref.~\cite{Zhang2024}, which serve as benchmarks, and their generalized and
hardness-dependent counterparts described in the previous sections. Note
that $a = 5.5$ for the benchmark protocols but $a = 5$ in our study. The
$\HP$-linear protocol (with six degrees of freedom) yields the best results
for all graphs, and in particular for the hard ones with a tenfold improvement
over the benchmark.

\begin{figure}[h]
	\includegraphics[width=0.953\columnwidth]{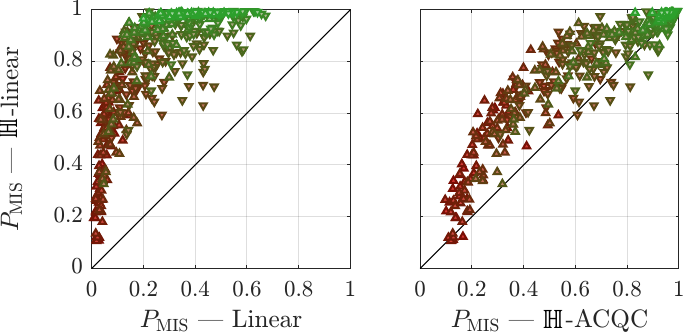}
	\vspace{-0.28em}
	\caption{Graph-to-graph comparisons of $\PMIS$. Left:~The
	$\HP$-linear protocol developed in this study leads to a
	3.6-fold mean improvement on the linear protocol described
	in Ref.~\cite{Zhang2024}. Right:~Except for rare instances,
	the hardness-dependent piecewise linear protocol always beats its
	ACQC counterpart.
	Note that $\PMIS$ depends strongly on $\HP$ but not on $N$ (upward
	or downward triangles), with low success probabilities for the
	hard graphs (maroon) and nearly perfect results for the easy ones (green).}
	\label{fig:optim_PMIS_compare}
\end{figure}

Graph-to-graph comparisons are illustrated in Fig.~\ref{fig:optim_PMIS_compare}.
As for the optimized schedule parameters, we observe that the values of
$\PMIS$ depend mostly on the hardness parameter and not on the size of the graphs.
For instance, the success probabilities for the $\HP$-linear
protocol range from 0.59 ($\HP = 4.20$) to 0.99 ($\HP = 0.46$) for $N = 10$,
and from 0.11 ($\HP = 9.71$) to 0.98 ($\HP = 0.52$) for $N = 17$.

\section{Cloud quantum computing demonstrations}
\label{sec:demonstrations}

On the basis of thorough numerical simulations, we showed in the previous section
that hardness-dependent schedules lead to high probabilities of reaching the MIS
solution. We first extend these results to the hardware and to graphs that are
not part of the dataset used in the optimization. First, we consider a small
set of ``toy graphs'' that we can still simulate classically to ensure that the
various protocols are correctly run in Aquila. Then we apply the models
to large graphs made up of 137 vertices to verify that the hardness parameter
and not the size of the graph is indeed the most relevant property when solving
the MIS problem.

For all examples, we use 720 shots (repetitions of the adiabatic evolution) to
get enough data for a statistical comparison of the methods. We choose this
specific value because it is a multiple of 9, 12, and 16, which are the maximum
number of copies of the small graphs ($3\times3$ and $3\times4$ lattices with
$a = 5$ and $a = 5.5$) that fit and can be run simultaneously in the hardware.
We consider only the shots in which the lattice is correctly filled by atoms.
The typical probability of failing to occupy a site in Aquila is $0.007$ (see
Sec. 1.5 in Ref.~\cite{Wurtz2023}), and in practice we get $684\pm 7$ and
$380\pm 14$ valid shots for the small and the large graphs, respectively.

\subsection{Toy graphs}
\label{sec:demonstrations_toys}

\begin{figure}[t]
	\includegraphics[width=0.972\columnwidth]{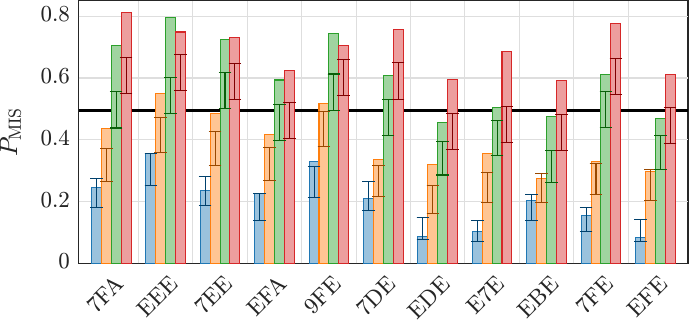}
	\caption{Probability of reaching the MIS solution for the 11 toy graphs. The
	rectangular bars are numerical results and the error bars are the results of
	the demonstration (95\% confidence intervals). The ranking of the four
	protocols studied is the same as in Fig.~\ref{fig:optim_PMIS_box}, with
	$\HP \in [2,3]$ here. The horizontal black line represents the average success
	probability obtained by the Bayesian optimization in Ref.~{Finzgar2024}
	(see Fig.~9 therein), which performs better than the benchmarks (blue and
	orange) but not as well as the hardness-dependent protocols (green and red).}
	\label{fig:demonstrations_toys}
\end{figure}

In Fig.~\ref{fig:demonstrations_toys}, we display the probability of reaching
the MIS solution for the 11 toy graphs defined in~\cite{Finzgar2024}. As in
that article, we set $\tf = 0.7$ so that the probabilities of success are
distributed around 50\%. We observe good agreement between numerical
simulations and the corresponding results obtained on hardware, especially for
the benchmark protocols. A possible reason for the optimized schedules
yielding outcomes that are slightly lower than expected may be the choice of a
smaller lattice spacing or the larger gradient of $\W$ at the end of the
evolution; many other sources of error could explain these
differences\textemdash see Sec. 1.5 in~\cite{Wurtz2023} for an exhaustive list.
Nevertheless, the $\HP$-linear protocol still performs better on the hardware
than the linear and CD benchmarks, with a threefold mean improvement and a
twofold mean improvement, respectively.

\subsection{Large graphs}
\label{sec:demonstrations_large}

\begin{figure}[h!]
	\includegraphics[width=0.972\columnwidth]{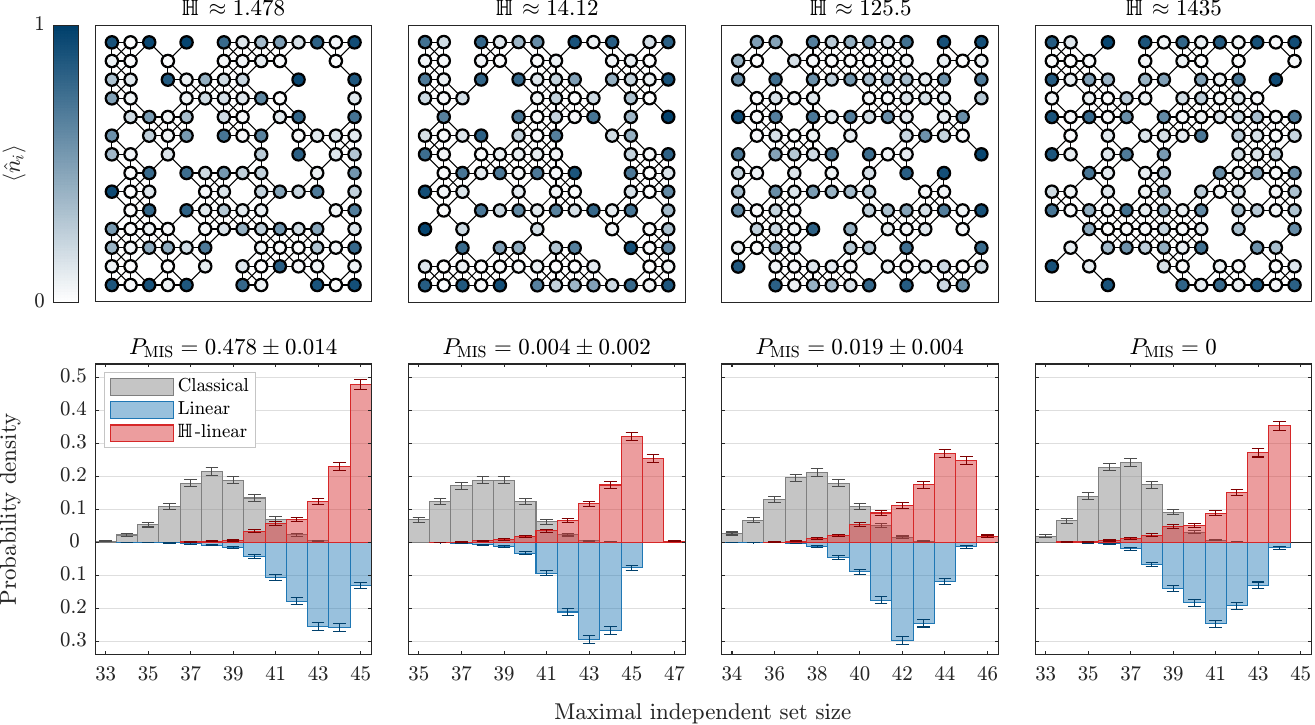}
	\caption{Demonstrations for large unit disk graphs.
	Top:~Four graphs with 137 vertices each and whose hardness parameters span
	4 orders of magnitude; the three hardest graphs are taken from
	\cite{Finzgar2024}. In shades of blue, the probability $\braket{\hat{n}_i}$
	to measure the $i$th atom in the Rydberg state $\ket{1}$ at the end
	of the $\HP$-linear protocol. Because of the high degeneracy of the MIS,
	some vertices are equally measured in the ground state or the excited state.
	Bottom:~Results of the demonstrations (after a minimal postprocessing)
	to find an independent set of a given size, with the MIS solution at the
	rightmost bin. The error bars represent the 95\% confidence intervals.
	The classical greedy algorithm (gray) fails to find the solution in all cases.
	The benchmark linear protocol (blue) reaches the MIS in the easiest case but
	fails for any harder instance. Finally, the $\HP$-linear protocol is able to
	find the MIS except for the hardest graph; the corresponding values of
	$\PMIS$ are indicated. The success probabilities for the two
	CD protocols (not shown) lie between those of the two linear protocols.\newline}
	\label{fig:demonstrations_large}
\end{figure}

We now turn to large graphs, which are of much greater interest in real applications.
We consider three of the four graphs studied in Ref.~\cite{Finzgar2024} (see
Fig.~10 therein), which consist of 137 vertices and whose hardness parameters extend over 3
orders of magnitude ($\HP \approx$ 14, 126, and 1435). In addition, we generate
one very easy graph of the same size to span yet another order of magnitude
($\HP \approx$ 1.4); see Fig.~\ref{fig:demonstrations_large}. In these examples,
the hardness parameter can still be computed by an advanced classical algorithm;
namely, by the GenericTensorNetworks \texttt{Julia} library~\cite{Liu2021,Liu2023}.
However, in general, this parameter cannot be considered to be known beforehand
as it depends on the MIS solution, which is the very target of the demonstrations.
Therefore, we apply the optimized schedules in the limit $\HP \to \infty$ in
what follows, and we focus on the piecewise linear protocols as they are expected
to yield the best results. 

Contrarily to the small graphs, we now use the full coherent time of the
hardware, i.e. $\tf = \tmax = 4$. We want the probability of finding a MIS
solution to be as high as possible for the hardest instances. Moreover, we
postprocess all data of the demonstrations such that the measured outcomes
become valid maximal independent sets. To this end, we use the greedy procedure
described in Sec. ~S2.3 in Ref.~\cite{Ebadi2022}: First, we remove any of two
neighboring vertices until the sets become independent. Second, we include a
vertex at any free site with no neighbor until the sets are maximal. The same
procedure applied to an empty configuration is used as a classical benchmark,
which highlights the relative gain obtained with the quantum protocols~\cite{Wurtz2023,
Wurtz2024b}.
 
The results are summarized in Fig.~\ref{fig:demonstrations_large}: The MIS
solution is never found by the classical greedy algorithm and is obtained by
the standard linear protocol only for the easiest graph. In contrast, the
$\HP$-linear protocol leads to the solution for all but the hardest graph,
which indicates a clear improvement over the existing protocols.

\section{Discussion and outlook}

In this work, we have shown how to design the adiabatic schedules of
the Rydberg Hamiltonian to efficiently solve the MIS problem on unit disk graphs.

First, a statistical investigation of the conditions for which the final
quantum state encodes the MIS solution has revealed that the final detuning should
lie in a narrow interval. To our knowledge, these are the tightest
bounds ever mentioned. We believe that choosing a final detuning in
that interval would improve the results of many other studies, either in terms
of success probability or in terms of the number of optimization steps. Indeed,
the application of adiabatic protocols in which the final ground state does not correspond
to the sought solution can only lead to poorly performing algorithms.

Second, the individual optimization of numerous small graphs has shown that the
optimum parameters of the schedules depend mostly on the hardness parameter
and only marginally on the size of the graph. Through extensive classical
simulations of small systems, we have generated adiabatic schedules
that perform extremely well also when applied to much larger graphs;
only the hardest instances are not solved by our protocol. Obviously,
improvements in the hardware would greatly help in this regard, especially
if they lead to a longer coherent evolution time. Since quantum technology
is evolving rapidly, this could be the case soon. However, because of the
combinatorial nature of the hardness parameter, very hard MIS problems exist
only in sufficiently large graphs, so our numerical study should be
extended to larger graphs to address this issue. We believe that improved
schedules could be easily found by consideration of slightly larger systems,
with up to about 25 vertices, or even larger ones with the use of tensor
network techniques for the simulations whenever appropriate~\cite{Orus2019}.

We have focused our work on two families of schedules; namely, piecewise
linear and smooth counterdiabatic functions. The phase of the
Rabi drive is nonconstant only in the latter case, and we have developed
a generalized version of the ACQC protocol~\cite{Zhang2024} to make it
specific to each graph. However, when correctly parametrized, the somehow
simpler $\HP$-linear protocol performs better in almost all numerical
simulations. This difference is even more pronounced in demonstrations,
as the smooth CD schedules have to be converted into piecewise linear or constant
functions. We have used a maximum number of six free parameters in the
optimization, but more general functions may be investigated since
the method described in this work applies to any kind of schedule.
Moreover, it would be interesting to develop models of optimized schedules
that do not depend only on the hardness parameter but that depend also on other global
properties of the graphs, e.g., the mean degree or the mean eccentricity
of the vertices. On the basis of machine learning, such additional predictors
could improve the fitting of the schedule parameters.

We think that hybrid classical-quantum algorithms could take advantage
of our protocols and, more generally, of precomputed optimized schedules:
instead of searching in the huge space of
all parameters, one could consider lower-dimensional meaningful ansatzes,
such as the $\HP$ parametrization of the schedules or the principal
component of the residuals of the fitted parameters.

Finally, we believe that the piecewise linear protocols developed in
this study should serve as a benchmark for any future adiabatic
algorithm that aims to solve the MIS problem with Rydberg atoms.
In particular, the most basic version of this protocol with only four free
parameters and in the limit of infinite hardness admits a very simple
mathematical description\textemdash see Fig.~\ref{fig:optim_schedules_Lin4_HP}\textemdash
and it leads to relatively good results regardless of the specific
graph to be solved.

\section*{Data availability}
The data that support the findings of this article are openly available~\cite{MyGitHub}.

\onecolumngrid
\noindent 
\twocolumngrid


\clearpage
\onecolumngrid

\appendix
\renewcommand{\appendixname}{APPENDIX}

\section{\MakeUppercase{Ground state encoding of the maximum independent set problem}}
\label{sec:MIS}

Given a simple undirected graph $\mathcal{G} = (V,E)$ with vertex set $V$ and edge set
$E$, a subset $S \subseteq V$ is called independent if no pair of vertices in
$S$ are connected by an edge in $E$. In the MIS problem,
we want to find the independent set of largest cardinality. If we denote by
$x_i \in \{0,1\}$ the absence or the presence of the vertex $i=1, \ldots, N$ in $S$,
it is proven that the MIS solution minimizes the cost function \cite{Boros2002,Choi2008}
\begin{equation}
	\mathcal{C}(x_1, \ldots, x_N) =  - \sum_{i\in V} \D\,x_i
	+ \sum_{ij\in E} V_{ij}\,x_i x_j
\end{equation}
if $0 < \D < V_{ij}$ for all $ij\in E$. In analog quantum computing with arrays of Rydberg atoms, $\mathcal{G}$ is a unit disk graph that connects the nearest and next-nearest neighbors
of a square lattice. If we ignore the interactions between farther atoms,
for a non-planar graph (i.e., when crossing diagonals belong to it),
the MIS condition becomes
\begin{equation}
	0 < \Df < \frac{C_6}{(\sqrt{2} a)^6} = \frac{V_0}{8},
	\label{eq:MIS-condition}
\end{equation}
where $a$ is the lattice spacing, $V_0 = C_6/a^6$ is the interaction strength between adjacent atoms, and $\Df = \D(\tf)$ is the detuning at the final evolution time $\tf$.
In several investigations, it is implicitly assumed that the
inequalities (\ref{eq:MIS-condition}) hold for all unit disk graphs due to
the sharp Rydberg blockade transition. However, in what follows we show that
the condition $0 < \Df$ is necessary but not sufficient.

\subsection{Nonexistence of a lower bound on the detuning}

\begin{figure}[h]
	\includegraphics[height=5.25cm]{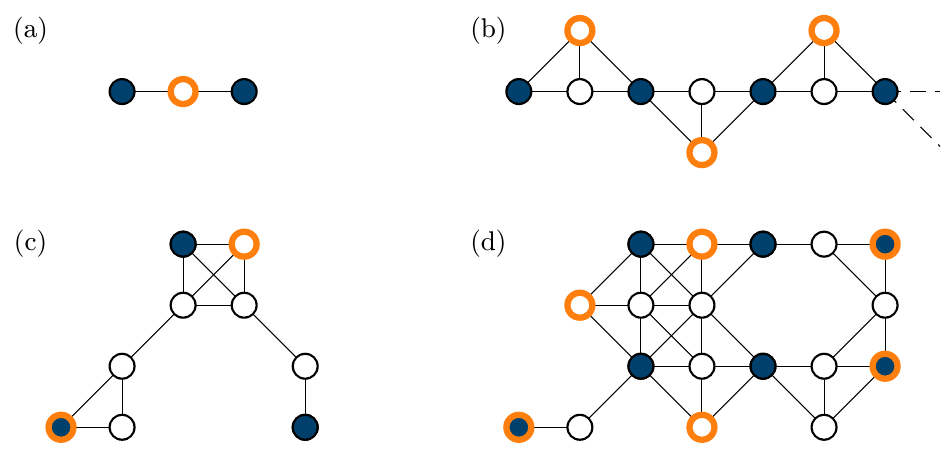}
	\caption{
	Examples of unit disk graphs to study the lower bound on the final
	detuning. The MIS solution (filled circles) has a lower energy than
	the suboptimal configuration (orange circles) if and only if $\Df > \DLB$.
	(a) Reference graph used in Eq.~(\ref{eq:DLB_trivial}).
	(b) Graph with $N=3m+1$ vertices forming $m$ alternating upward and
	downward triangles. For any $\Df > 0$, there exists a value of $m$ such
	that the suboptimal configuration is energetically favorable.
	(c), (d) Two non-planar graphs with very different lower bounds:
	(c) $\DLB \approx V_0/1017$ ($\HP \approx 0.42$) and
	(d) $\DLB \approx V_0/11.49$ ($\HP = 15$).}
	\label{fig:bounds_examples}
\end{figure}

The van der Waals interaction decays quickly with the distance between atoms,
but it cannot be ignored for distances close to the blockade radius. As an
example, consider the graph that consists of three atoms in a row with distance
$2a$ between its extremities; see Fig.~\ref{fig:bounds_examples}(a). In this
case, the MIS solution is $(1,0,1)$, which corresponds to an energy of
$-2\Df+V_0/64$. By listing the energy of all other configurations, we find the
following conditions on $\Df$ such that $\ket{101}$ is indeed the ground state
of the final Rydberg Hamiltonian:
\begin{equation}
	V_0/64 < \Df < 2\,V_0.
	\label{eq:DLB_trivial}
\end{equation}
The upper bound is, of course, already satisfied in Eq.~(\ref{eq:MIS-condition})
as it is a sufficient condition, but the value $V_0/64$ should be considered
as a strict lower bound for any protocol; note that this bound is mentioned in
Ref.~\cite{Wurtz2023}.
The question is then to determine if there exists, for unit disk graphs
in a square lattice, a finite lower bound on the detuning to ensure the
equivalence between the classical MIS solution and the ground state of
the quantum Hamiltonian. Unfortunately, the answer is negative.
Consider the unit disk graph in Fig.~\ref{fig:bounds_examples}(b),
and the following two sets of vertices: $S_0$, the MIS solution with
$m+1$ vertices on the main line (filled circles); $S_1$, the $m$ vertices
located at the tip of the alternating upward and downward triangles (orange circles).
If we take into account all pairwise interactions, the corresponding energies read
\begin{subequations}
	\begin{align}
		E_0 &= -(m+1)\,\Df + \frac{V_0}{64}\sum_{i=1}^m \frac{m+1-i}{i^6},
		\label{eq:MIS-energies-A}\\
		E_1 &= -m\,\Df + \frac{V_0}{64}\sum_{i=1}^{\lfloor m/2 \rfloor} \left(
		\frac{m+1-2i}{(4i^2-4i+2)^3} + 
		\frac{m-2i}{(2i)^6}\right).
		\label{eq:MIS-energies-B}
	\end{align}
\end{subequations}
For the ground state of the Hamiltonian to yield the MIS solution we must have
$E_0 < E_1$, i.e., the detuning has to satisfy the inequality
\begin{equation}
	\frac{V_0}{64} \left( \Sigma_0-\Sigma_1\right) < \Df,
\end{equation}
where $\Sigma_0$ and $\Sigma_1$ stand for the sums on the right-hand sides of
Eqs.~(\ref{eq:MIS-energies-A}) and (\ref{eq:MIS-energies-B}).
However, one can check that the difference $\Sigma_0-\Sigma_1$ is always positive
and becomes arbitrarily large for increasing $m$.
In particular, this difference is larger than 8 for $m \geq 9$,
leading to a contradiction with Eq.~(\ref{eq:MIS-condition}).

In conclusion, there does not exist a value of the final detuning
such that it is ensured that the Rydberg Hamiltonian encodes the MIS solution
of all unit disk graphs.

\subsection{Statistical analysis of the bounds on the detuning}
\label{sec:MIS_bounds}

\begin{figure}[t]
	\includegraphics[height=6.5cm]{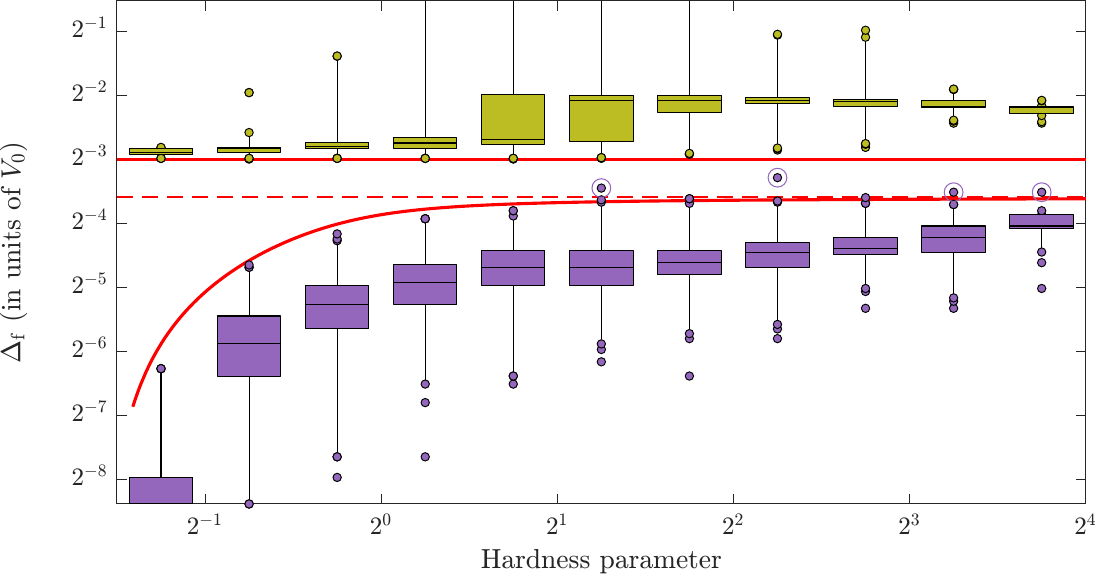}
	\caption{Box plot of the lower bounds $\DLB$ (purple) and upper bounds $\DUB$
	(yellow-green) on the detuning $\Df$ such that the ground state of the
	Hamiltonian encodes the MIS solution. Calculations are performed on $10\,000$
	nonisomorphic unit disk graphs with 8\textendash 19 vertices. All but four
	graphs satisfy $\DLB \leq V_0/12 $ (dashed red line) while $\DUB$ is always
	strictly larger than $V_0/8$. It is interesting to note that, on average,
	easy graphs exhibit much	smaller lower bounds than hard graphs (red curve);
	see	also Fig.~\ref{fig:bounds_examples}(c-d). Since the average value of
	$\DLB$ quickly saturates for not-too-easy graphs, we are confident that these
	bounds can be effectively applied to graphs of any size.}
	\label{fig:bounds_analysis}
\end{figure}

As shown above, the final detuning has to be arbitrarily high if one wants
to encode the MIS solution of any graph in the Rydberg Hamiltonian. However,
this is not the case in general, as most graphs obey well-defined conditions
on $\Df$. From an extensive analysis of $10\,000$ graphs\textemdash see
Fig.~\ref{fig:bounds_analysis}\textemdash we find that more than 99.9\% of all graphs
satisfy the following bounds:
\begin{equation}
	\Df \in [\DLB, \DUB] =
	\left[ \frac{V_0}{12} \,, \frac{V_0}{8} \right].
	\label{eq:mis_detuning_range}
\end{equation}
Interestingly, this interval is very close to the best range of parameters
found by a variational optimization (see Fig.~6.1 in Ref.~\cite{Wurtz2023}) with
$\DLB = 28.91$\,MHz and $\DUB = 43.36$\,MHz for QuEra's Aquila device and
$a = 5$\,\textmu m. Note that a statistical analysis is also performed
in Ref.~\cite{Ebadi2022} (see Sec.~S2 therein), where it is stated that about 99\% of the MIS
problems can be encoded in the ground state of a Rydberg Hamiltonian for
well-chosen fixed parameters of the quantum hardware. We strongly recommend
considering the interval in Eq.~(\ref{eq:mis_detuning_range}) when one is setting the
final detuning of any adiabatic protocol for the MIS problem. A value that lies
far outside this interval will likely lead to poor results simply because the
ground state of the Hamiltonian does not correspond to the MIS solution.

\section{\MakeUppercase{Graph-dependent counterdiabatic driving}}
\label{sec:CD}

For convenience, we set $\hbar = 1$ and $\phi(t) = 0$ in what follows, but all results can be extended to the general scenario. In this case, the Rydberg Hamiltonian reads
\begin{equation}
	H(t) = \frac{\W(t)}{2}\sum_i \X_i - \D(t) \sum_i \n_i +
	\sum_{i<j}V_{ij}\n_i\n_j \equiv
	\frac{\W(t)}{2}\ \X - \D(t)\ \n + \V,
	\label{eq:phase_free_hamiltonian}
\end{equation}
where $\n_i = \ketbra[i]{1}{1}$ and $\X_i$ is the Pauli matrix
$\ketbra[i]{0}{1} + \ketbra[i]{1}{0}$ acting on atom $i=1, \dotsc, N$.

\subsection{Approximate adiabatic gauge potential}

The idea of the CD driving is to add a gauge potential $\A$ to the Hamiltonian
$H$ such that the transitions between eigenstates are suppressed during
a shortened adiabatic evolution:
\begin{equation}
	\widetilde{H}(t) = H(t) + \A(t),
	\label{eq:HCD}
\end{equation}
where $\A$ satisfies the equation [see Eq.~(2) in Ref.~\cite{Sels2017}]:
\begin{equation}
	\left[ \ic \partial_t H - \left[\A, H \right], H\right] = 0.
\end{equation}
If the interactions $\V$ are ignored, it is shown in Ref.~\cite{Zhang2024} that a
solution of this equation exists:
\begin{equation}
	\A_{V\rightarrow 0} = -\frac{1}{2}\ \frac{\W\dot{\D} - \dot{\W}\D}
	{\W^2+\D^2}\ \Y.
\end{equation}
This gauge potential leads to improved results compared with a simple adiabatic protocol,
but one might expect even better results with a gauge potential that explicitly depends
on the coefficients $V_{ij}$, hence being specific to the graph to be solved.
To find such a generalized gauge potential, we consider the variational
procedure that consists in minimizing the action [see Eq.~(8) in Ref.~\cite{Sels2017}]
\begin{equation}
	\mathcal{S}(\A_{\lambda}) = \Tr\left( G^2(\A_{\lambda}) \right)
	\text{, with }
	G(\A_{\lambda}) = \partial_t H + \ic [\A_\lambda, H].
\end{equation}
Here we use the ansatz $\A_\lambda = -\Y \lambda(t)/2$ since the one-body
Pauli matrices $\Y_i$ are the only operators that contribute to the action
while being implementable in the quantum hardware through the phase $\phi$ of
the Rabi drive. One can easily check that $\mathcal{S}$ is quadratic in $\lambda$
and that the optimum value of this parameter is given by
\begin{equation}
	\lambda = 2\,\frac{\Tr\left(\partial_t H\, [\ic\Y,H]\right)}{\Tr\left([\ic \Y,H]^2\right)}.
\end{equation}
On the basis of the algebraic properties of the Pauli matrices and the trace operator,
this expression can be developed and simplified as follows:
\begin{equation}
	\lambda = \frac{\W\dot{\D} - \dot{\W}\D + \dot{\W}\ T_1}
	{\W^2 + \D^2 - 2\D\ T_1 + T_2},
	\label{eq:lambdaCD}
\end{equation}
where the normalized traces $T_1$ and $T_2$ are given by
\begin{subequations}
\label{eq:traces}
\begin{align}
	N\, T_1 &= 2^{-N} \Tr\left(\X\, [\ic \Y,\V]\right) = \sum_{i<j}V_{ij},			
	\label{eq:T1}\\
	N\, T_2 &= 2^{-N} \Tr\left([\ic \Y,\V]^2\right) =
	\frac{1}{2}\sum_{i\leq j} \sum_{k \neq i,j} V_{ik}V_{kj}.
	\label{eq:T2}
\end{align}
\end{subequations}
As expected, the graph-dependent gauge potential $\A_\lambda$ is equal to
$\A_{V\rightarrow 0}$ in the case of vanishing interactions, with $T_1=T_2=0$
in this limit.
Because of the fast-decaying interaction between distant atoms, $T_1$ and $T_2$ do
not scale with $N$ as they represent average values of (nearly) local functions.
For instance, $T_1$ is exactly half the mean degree of the underlying graph if one
considers its adjacency matrix instead of the interaction terms in Eq.~(\ref{eq:T1}).

Finally, the driving amplitude and phase of the full Rydberg Hamiltonian
$\widetilde{H}$ become
\begin{subequations}
\label{eq:WPCD}
\begin{gather}
	\widetilde{\W} = \sqrt{\W^2 + \lambda^2},\label{eq:WCD}\\
	\widetilde{\phi} = \arctan(\lambda/\W).\label{eq:PCD}
\end{gather}
\end{subequations}
Note that $\lambda$ and therefore $\dot{\W}$ should be equal to zero at the beginning
and at the end of the protocol to satisfy the boundary conditions
$\widetilde{\W}(\ti) = \widetilde{\W}(\tf) = 0$.

\subsection{Practical application of the counterdiabatic driving}

\subsubsection{Hardware implementation}
It follows from Eq.~(\ref{eq:WCD}) that $\widetilde{\W}$ is always larger than $\W$,
so it may exceed the maximum Rabi frequency $\Wmax$
of the quantum computer. In this case, we rescale the adiabatic function $\W$
by a factor $0 < \kappa < 1$ to satisfy the hardware constraint at any time $t$ of the
evolution:
\begin{equation*}
	\max_t \left\{ \widetilde{\W}\big(\kappa\,\W(t), \D, T_1, T_2\big)\right\} = \Wmax. 
\end{equation*}
Note that this is slightly different from rescaling $\widetilde{\W}$ directly,
as the relations (\ref{eq:lambdaCD}), (\ref{eq:WCD}), and (\ref{eq:PCD}) would
not hold in that case.

\subsubsection{Variational parameter}
\label{sec:CD_variational}
Since Eq.~(\ref{eq:lambdaCD}) is a generalization of the counterdiabatic term proposed
in Ref.~\cite{Zhang2024} through the addition of $T_1$ and $T_2$, it is interesting
in an optimization context to consider a continuous interpolation between these two protocols.
One possibility is to consider the interaction strength $V_0$ as a free parameter
in the range $[0, C_6/a^6]$ when one is calculating the CD driving. To avoid any confusion
with the fixed interactions in the Rydberg Hamiltonian, we denote by $\nu$ this free
parameter. Numerically, this is equivalent to replacing $T_1$ and $T_2$ in Eq.~(\ref{eq:lambdaCD})
by the following expressions:
\begin{subequations}
	\label{eq:T_nu0}
	\begin{align}
		T_1 &= \nu \, T_1^{(0)},\\
		T_2 &= \nu^2 \, T_2^{(0)},
	\end{align}
\end{subequations}		
where $T_1^{(0)}$ and $T_2^{(0)}$ are the traces evaluated at $V_0 = 1$ in
Eqs.~(\ref{eq:T1}) and (\ref{eq:T2}). Another possibility would be to consider the
two-variable parametrization $T_i = \nu_i\,T_i^{(0)}$, with
$\nu_i \in [0, V_0^i]$ and $i\in\{1,2\}$.

\section{\MakeUppercase{Graph datasets and numerical implementation}}
\label{sec:datasets}

In the following sections, we first describe two sets of unit disk graphs $\mathcal{G} = (V,E)$
that are used in this study; the data are available on GitHub \cite{MyGitHub}.
In all examples, we consider an underlying
square lattice where the distance $a$ between nearest-neighbors is $1$ and
where any pair of vertices in $V$ is connected by an edge in $E$ if the 
corresponding distance is smaller than or equal to $\sqrt{2}$. Then we detail
the numerical method that is used to simulate the evolution of the quantum
state under the action of the Rydberg Hamiltonian.

\subsection{Graph datasets}
\label{sec:datasets_graphs}

\paragraph*{Small graphs}

\begin{figure}[h]
	\includegraphics[width=0.92\columnwidth]{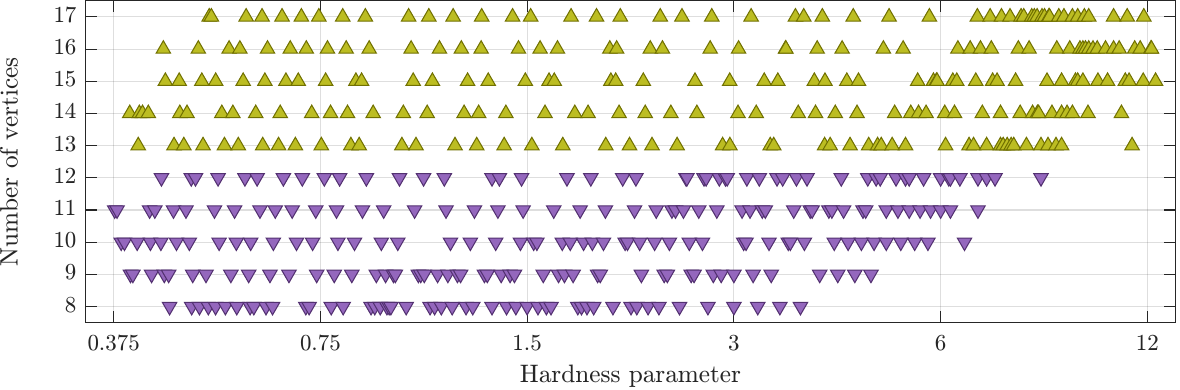}
	\caption{Visualization of the hardness parameters $\HP$ and the orders of the 500
	small graphs. Because of the combinatorial definition of the hardness parameter,
	graphs with few vertices (6--12 purple triangles) cannot generate very hard MIS
	instances. This explains why larger graphs (13--17 vertices, yellow-green triangles)
	are overrepresented in the range $\HP \in [6,12]$.}
	\label{fig:dataset_small_hp}
\end{figure}

From $10\,000$ nonisomorphic graphs randomly generated, we choose 500 graphs
such that the correlation between their hardness parameter and their order is as
small as possible; see Fig.~\ref{fig:dataset_small_hp}. More precisely, we apply
the following rules to determine which graphs belong to the dataset:
\begin{compactitem}
	\item There are exactly 50 graphs for each order from 8 to 17.
	\item The hardness parameters are distributed as evenly
	as possible on a logarithmic scale from 0.375 to 12; in particular,
	no two graphs share the same hardness parameter, and exactly 100 graphs
	are contained in each bin $[3\times 2^{i-4}, 3\times 2^{i-3})$ with $i = 1, \ldots, 5$.
	\item The distribution of the graph orders is as uniform as possible
	within each such bin.
\end{compactitem}

\paragraph*{Toy graphs}
For completeness, we display in Fig.~\ref{fig:dataset_toys} the 11 toy graphs
defined in Appendix C in ~\cite{Finzgar2024}:
\begin{compactitem}
	\item The underlying square lattice has $4\times3$ sites.
	\item The order of the graphs is 9 or 10.
	\item There is only one MIS solution for each graph.
\end{compactitem}

\begin{figure}[h]
	\includegraphics[height=7cm]{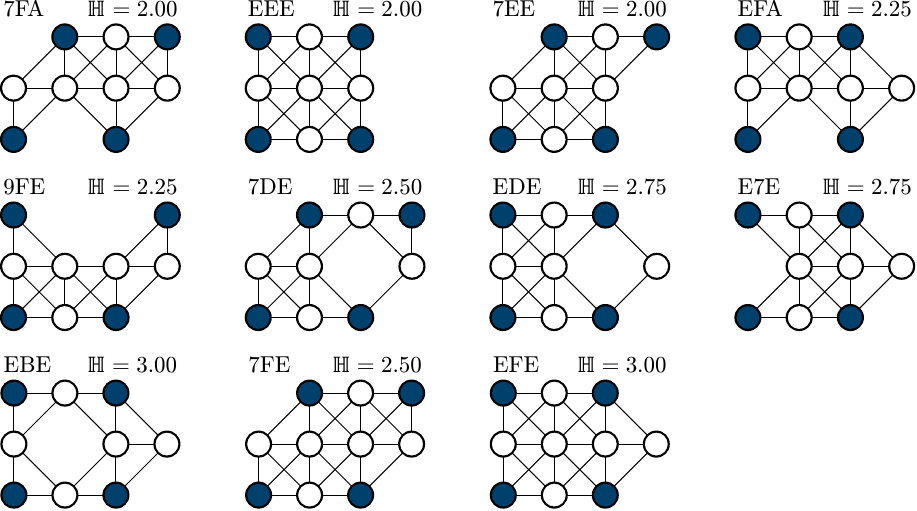}
	\caption{The 11 toy graphs sorted by the number of vertices (nine or ten)
	and by hardness parameter (between 2 and 3). The unique MIS is highlighted with
	dark-blue circles. To refer to a specific graph, we generate the
	three-nibble identifiers as follows: the $i$th nibble (4-bit aggregation)
	is the hexadecimal symbol that represents the absence or presence
	of a vertex in the four sites of the $i$th row (from top to bottom).}
	\label{fig:dataset_toys}
\end{figure}

\subsection{Numerical implementation}
\label{sec:datasets_implementation}

Several software development kits exist to simulate the Rydberg evolution: for example,
\texttt{BLOQADE} \cite{Bloqade2024} and \texttt{AMAZON BRAKET} \cite{AWS2024} for
QuEra's hardware, or \texttt{PULSER} \cite{Silverio2022} based on QuTiP for Pasqal's technology.
Nevertheless, we developed our own routines to get full control over all
parameters and, more importantly, to fit exactly our needs. Optimization of
hundreds of graphs for several protocols requires intense (classical) computing
power, so we have to find the perfect balance between accuracy and speed.
For example, for each graph of the datasets we precompute all time- and
protocol-independent terms in the Hamiltonian, namely, the $\hat{X}$, $\hat{Y}$,
$\hat{n}$, and $\hat{V}$ operators; see Eq.~(\ref{eq:phase_free_hamiltonian}).
We also use an explicit Runge-Kutta method of order 2 \cite{Bogaki1989} to
integrate the time-dependent Schr{\"o}dinger equation, which turns out to be
sufficiently accurate and much faster than the default order 8 method
(\texttt{DOP853} in \texttt{PYTHON} or \texttt{VERN8} in \texttt{JULIA}). The initial
state of this ordinary differential equation is $\ket{0}^{\otimes N}$,
which corresponds to the unique ground state of the Rydberg Hamiltonian
with $\W = 0$ and $\D < 0$.

\subsubsection{Blockade subspace}
\begin{figure}[t]
	\includegraphics[width=0.875\columnwidth]{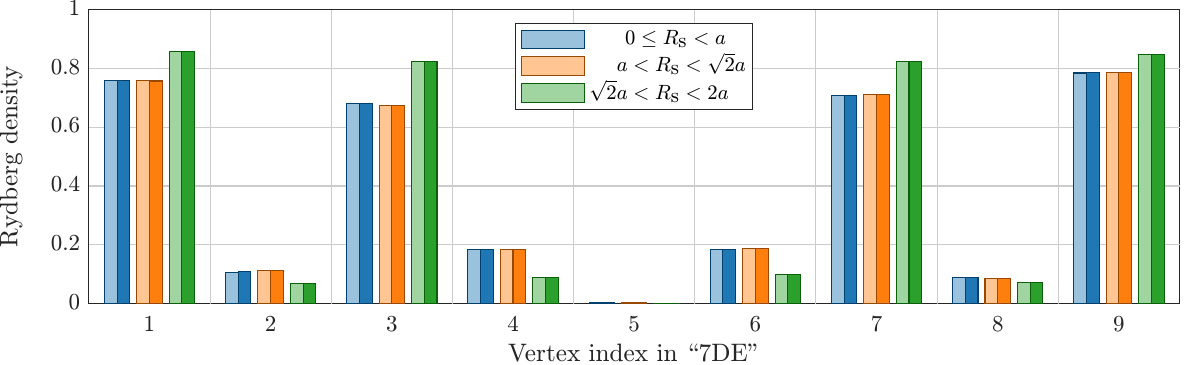}
	\caption{Numerical simulations of the Rydberg Hamiltonian for the toy
	graph ``7DE'' and the ACQC protocol defined in Ref.~\cite{Zhang2024} with
	$\tf = 1$\,\us. The height of the bars represents the probability to
	find the atom $i$ in the state $\ket{1}$ at the end of the evolution,
	with $i = 1, \ldots, 9$ from top to bottom and left to right in the graph.
	The MIS solution (1, 3, 7, 9) is clearly visible in this bar chart. 
	We verify that our implementation (light colors) is perfectly equivalent
	to that of \texttt{BLOQUADE} (dark colors); note that, depending on the
	software development kit, the
	phase has to be multiplied by minus one in Eq.~(\ref{eq:PCD}).
	We also observe that the simulations in the full Hilbert space (blue bars)
	match those performed in the blockade subspace of nearest neighbors (orange bars).
	However, the unit disk subspace approximation $\Rs = \Rb$ (green bars)
	is clearly in favor of the MIS solution and thus cannot be considered as
	satisfactory. This is even more noticeable when comparing the
	probability to reach the ground state at the end of the evolution for
	each blockade subspace: 0.49(4), 0.49(6), and 0.70(0) respectively.}
	\label{fig:numerical_subspace}
\end{figure}

Besides the above-mentioned implementation aspects, a crucial ingredient is the
correct use of the blockade subspace: when the interaction
energy between Rydberg states is much larger than the Rabi strength, the
exact dynamics in the full Hilbert space is well approximated by the dynamics
in the subspace where only one Rydberg excitation is allowed between
nearby atoms \cite{BloqadeSubspace2024}. In numerical simulations, the
states $\ket{11}_{ij}$ are therefore ignored if the distance between
$i$ and $j$ is smaller than a given subspace radius $\Rs$, which results in
much faster computations. The default setting in most programs
is $\Rs = \Rb$, so the quantum evolution is restricted to the independent sets
in unit disk graphs. However, this approximation is too coarse to get accurate
results\textemdash see Fig.~\ref{fig:numerical_subspace}: ignoring the states $\ket{11}$
for next-nearest neighbors (extremities of a diagonal) leads to success
probabilities that are overestimated by up to 50\%. Therefore, it is crucial
to simulate either the full Hilbert space or the blockade subspace
where only nearest-neighbor Rydberg states are ignored.

\subsubsection{Optimization algorithm}
In this study, we focus on the probability of reaching a MIS, i.e., we want
to maximize the overlap between the quantum state at the end of the evolution
and all MIS solutions of the graph under investigation; see Eq.~(\ref{eq:PMIS}).
The set of parameters $\Theta$ to be optimized depends on the protocol, e.g.,
$\Theta = (\taui, \tauf, \Di, \Df)$ for the simple linear schedules.
Since the Schr{\"o}dinger evolution is deterministic and we simulate
noiseless systems, any standard optimization algorithm may be applied.
We choose the simplex search method \cite{Lagarias1998},
which turns out to be robust and very efficient in our setting.
Finally, we set $\tf = 1$\,\us, which leads to success probabilities that are well
distributed over the whole unit interval for the 500 graphs of the dataset.

\clearpage
\twocolumngrid

\bibliography{ref}

\end{document}